Bohr meets Rovelli: a dispositionalist account of the quantum limits of knowledge

Mauro Dorato
Department of Philosophy, Communication and Media Studies
Università degli Studi Roma Tre,
Via Ostiense 234, 00146, Rome, Italy

1) Introduction: ignoramus and ignorabimus?[1]

In the last couple of centuries, the history of physics has at times been sparkled by two opposite attitudes, which to simplify we could denote a radically pessimistic and an overoptimistic one. As a famous representative of the former first camp, we find the German physiologist Emile Du Bois-Reymond, who in a famous speech held in 1880 at the Berlin Academy of science claimed that there were seven world enigmas (*Die Sieben Welträtzel*) that neither science nor philosophy could ever solve. Among these mysteries, he listed the nature of matter and force, the origin of life, the origin of intelligent thought and the question of free will.

As a representative of the overoptimistic camp, we can enlist Lord Kelvin when, a couple of decades later (April 27, 1900) gave a speech at the Royal Institute entitled "Nineteenth-Century Clouds over the Dynamical Theory of Heat and Light", in which he defended the view that physical knowledge was *complete*, so that the only future task for physicists was the "minor" one of providing more precise experimental measures of already known quantities. However, more realistically, deeply cognizant as he was of contemporary physical theories, mentioned "two clouds" - that as we now know, will lead to a thunderstorm! - namely the failure of Michelson and Morley's experiment and the ultraviolet catastrophe (clouds that have been swept away thanks to the special theory of relativity and to quantum mechanics respectively). By stressing the impassable limits of science, Du Bois-Reymond's Latin motto stirred a strong, optimistic reaction also on the part of the mathematical community. For example, in his 1930 speech at the Society of German Scientists, the great German mathematician David Hilbert expressed the view that: "In opposition to the foolish ignorabimus, our slogan shall be: Wir müssen wissen – wir werden wissen".[2]

---

[1] We don't know, we will never know
[2] "We must know, we will know" is the motto written on Hilbert's tomb.

Advices like that received at the end of the 19$^{th}$ century by the German physicist Max Planck are difficult to classify: older physicists discouraged him from studying physics, because according to these them there was nothing left to be discovered! Even though this counsel might have been motivated by the pessimistic outlook that physics was complete simply because it had reached the limits of the human knowledge, it should be clear why the optimistic motivation was also present, given the presumption of completeness reached by the contemporary physics. Relatedly, in 1981, in another famous essay that was certainly inspired by an overoptimistic attitude, the English physicist Steven Hawking claimed that the End of Physics might be in sight (1981), so that the major problems of physics would have been solved by the end of last century: "By this I mean that we might have a complete, consistent and unified theory of the physical interactions which would describe all possible observations" (Hawking 1981, p.15).

2   Quantum limits of knowledge: from sociology, to mathematics, to physics

Going now to our topic, we should keep in mind that there have various attempts to exploit quantum mechanics to claim that the theory has set an impassable limit to our capacity to know the natural world. The first explanations are mainly sociological or historical, and certainly more superficial than those arguments coming from mathematics and physics that will be presented subsequently.

2.1 Historical and sociological hypothesis

Some historians of quantum mechanics - like Forman (1984) - have very controversially claimed that a merely "instrumentalist" *interpretation* of the formulas of the emergent quantum mechanics (Bohr's *Copenhagen interpretation* in particular)[3] was influenced by the postwar revolt that characterized the Weimar republic and that contrasted in general rationalistic and realistic philosophies of science. According to Forman, in that cultural milieu the *scientific* fact that positions and momentum cannot be simultaneously known was regarded as evidence for a failure of causality, and therefore for the impossibility of a precise and exact knowledge of the physical world. To the extent that Bohr's hegemonic interpretation of quantum mechanics (that will be discussed below) is regarded as stressing too much the unknowability of certain aspect of nature, attempts to explain his and textbook views of quantum mechanics as due to mere *historical contingencies* have generated an interesting literature that here we will not review. The question posed in this historical and sociological context is: why formulations of the theory (for instance Bohm's) that are regarded as clearer and less mysterious (as well as empirically equivalent) were neglected until the 50ies even if they were available since the mid-twenties with De Broglie? (Cushing 1994).[4]

---

[3] We should keep in mind that the so-called Copenhagen interpretation has been regarded as an historical myth (2004).

Rather than discussing possible explananda of the hypothesis that quantum mechanics sets limit to our knowledge sociological theories that are *external* to physics proper, in what follows I will concentrate on the *internal* logical and conceptual aspects of quantum mechanics itself. In particular, in the next section I will sketch very important mathematical results (no-go theorems") that depend purely on the formal structure of the Hilbert space and I will then show the consequences that they have on the conceptual and experimental practice. In both cases, we will see how the ways in which such theorems are typically used to argue in favor of the existence of impassable limits of our knowledge of the physical world depend to an important extent on implicit philosophical stances about the crucial question: *what is quantum theory about?* (Laudisa 2014)

Rather than limiting in principle our knowledge of the external world, the structure of the Hilbert space poses some constraints on the ontology and metaphysics of quantum theory that certainly differ from those of classical physics and that are independently supported by the experimental practice and by a sound philosophical analysis. In a nutshell, my thesis is that both senses in which quantum mechanics can be legitimately regarded as a theory that sets important limits to our knowledge of the natural world (the formal and the experimental/philosophical one) that however *don't* justify Du-Boys Reymond's radically skeptical attitude to quantum mechanics

2.2 The mathematical no-go theorems

As in the case of the impossibility theorems demonstrated by Gödel, the mathematical structure of the Hilbert space by itself can be used to rule out in a rigorous way certain natural, commonsensical as well as classical assumptions about the physical world, specifically the *non-contextuality of possessed properties* (Kochen and Specker 1967).[5] After briefly revising two of the most significant no-go theorems, in the next section I will show in what sense the philosophical contents of the formal results were "anticipated" by Bohr's philosophical and conceptual analysis, which insisted on the very general fact that - except after having performed a measurement in which the state of the system is an eigenstate of the observable - *quantum measurements do not reveal preexisting values, but only values that we obtain after measurement*. Heisenberg's uncertainty relations are a particular case of this very general, characterizing feature of the theory and this simple truth is recognized also by defenders of Bohmian mechanics (Allori and Zanghì 2005), who press for the definiteness or non-contextuality of the positions of all the particles that compose all physical systems.

---

[4] What must be excluded is that quantum theory can be used to motivate certain irrationalistic attitudes that emerge from other corners of our culture: a superficial identification of indeterminism with irrationality neglects completely the fact that the main evolution equation in non-relativistic quantum mechanics, namely Schrödinger's, is deterministic and that Born's rule yields very precise even if merely probabilistic predictions that cannot be used to justify the existence of principled limits to our knowledge.

[5] Ghirardi's 1985, Isham's 1995, Hughes'1989 and Held's 2000 are good, accessible treatment of the problem.

The first important no-go results in the history of quantum mechanics is von Neumann's argument against the possibility of a hidden variables theory that be empirically equivalent to standard quantum theory (von Neumann 1955). These variables would assign a definite value to all the magnitudes describing a system. Already in this first example we will see the role of the philosophical intuitions in guiding the efforts of scientists and mathematicians.

By following the notation by Ghirardi (1985) let us indicate a hidden variable with the symbol H and three observables with the symbol A, B and C. Within a hidden variable theory, if we assume that A is a function of H, "the value A(H) is definite and must coincide with one of the values predicted by the standard theory" (Ghirardi 1985, p.178 of the Italian edition). Now suppose that the observable A is a linear combination of the other two observables B and C. If A is self-adjoint, then also any linear combination of A of B and C with numbers b and c must self-adjoint:

(i)   A=bB+cC

Von Neumann assumed that also the *definite values* assumed by the other two functions of the hidden variable, namely bB(H)+cC(H), must exemplify the same relation:

(ii)   A(H)=bB(H)+cC(H)

It can be shown, however, that (ii) is false because it does not hold generally (Bell 1966, p. 4) since it holds only for classical observables and for commuting observables. If the observables in questions are not simultaneously measurable or mutually compatible, then the determined values of two non-commuting observables do not satisfy (ii), but only their averages do. In a word, if the choice of *A*, *B*, *C*, is such that any two of them are incompatible, i.e. are not jointly observable, then von Neumann's result is unsuccessful, because in order to measure two incompatible observables one needs two different and incompatible measurement context or apparata.

Kochen and Specker's theorem (1967) presupposes a Hilbert space with dimension ≥3 but reinforces the case against the existence of hidden variables by assuming that the above mentioned three observables are pairwise compatible: A is compatible with B and with C, but B is not compatible with C and likewise for the other combinations. The theorem succeeds in showing that the actual formalism of quantum theory allows us to prove formally that a quantum system cannot be assigned "too many definite, intrinsically possessed magnitudes" (Isham 1995, p.199).

As is common to do in the literature, let us call the principle according to which any quantum system has definite properties at all times *value definiteness*. Kochen and Specker's theorem states that

(i)  if the state spaces has dimensions > 2 and

(ii) if B and C are incompatible observables,

then

if, one measures the observable A together with B one gets a different result than what one gets when A is measured together with C, observables. The result is of exceptional importance and has been demonstrated by using different techniques (see Held 2000). Its philosophical significance can be more simply be stated by claiming that it is the measurement context that defines the properties of physical system, so that the former and the latter constitute an unbreakable whole, something that, as we will see in the remainders of the paper, Bohr had very often insisted upon well before 1967.

### 2.3 The philosophical meaning of the no-go theorems and the quantum limits of knowledge

There are various revolutionary philosophical conclusions stemming from this theorem. If a realist about possessed properties wants to escape the strictures of the theorem by holding to the principle of value definiteness as much as possible, she must claim that *some* of them are definite and always possessed, while others are not. As is well known, in Bohmian mechanics, for instance, positions are always and everywhere possessed and definite, since they are the only non-contextual properties. All the other properties (like spin), on the contrary, also in Bohmian mechanics are contextual (see Clifton and Pagonis 1995). The significance of this theorem, therefore, is relevant also to those interpretations of quantum mechanics that reject pragmatist or instrumentalist views of the theory.

More in general, if the theorem "forces" us to change our conception of reality, in the sense that a physical system possesses *some* of its properties A (the state dependent ones) only in a relational or contextual sense, we cannot help but noticing that Bohr's interpretation of quantum mechanics were already explicitly in favor of the idea that finding out about the properties of an observable system calls for a specific measurement setting, which in its turn calls for an explicit kind of holism. As Bohr clearly stated in the 1949 volume in honor of Einstein, we must accept the "*impossibility of any sharp separation*

*between the behavior of atomic objects and the interaction with the measuring instruments which serve to define the conditions under which the phenomena appear"* (Bohr 1949, 210, emphasis in the original).

At this point we should add an important caveat. We should not ask from realist understandings of quantum mechanics conditions that the theory cannot satisfy. The question of the "knowability of the physical world" must always be asked within a particular physical theory and not in general. If intrinsically or contextually possessed properties of a physical system were a necessary condition for some form of realism about these properties, then we would have to conclude that quantum mechanics must be interpreted as an instrumentalist recipe for predictions. After all, in the macroworld, we don't believe that the properties of my table (like its shape or hardness) need a specification of "*the conditions under which the phenomena appear"*

Consequently, the question of the "quantum limits of knowledge" must be examined by presupposing the mathematical constraints of the theory, that reflect on a formal level those practical and experimental settings that are needed to come to know the properties of physical systems. In particular, we will discuss some aspects of this new relationist and contextualist conception of reality by comparing, in their apparent diversity, Bohr's holistic and Rovelli's relationist interpretation of the formalism, that deep down share a unifying metaphysics of dispositions and propensities. In a different sense also an aspect of Everettian quantum mechanics (1956,1957) can be interpreted as advocating a perspectival view of reality, in the sense that a perspective-free view of reality is impossible.

3   Bohr's ontological dispositionalism: the quantum unknowability of intrinsic properties

In order to set the framework of our discussion, we should avail ourselves of the distinction among intrinsic and extrinsic properties that has been defended by the great American philosopher David Lewis (1983). Even if this distinction is rather controversial and has been widely discussed in the metaphysical literature (see Marshall and Weatherson 2018 for a review) it will be prove illuminating for making sense of some of the aspects of the philosophy of quantum mechanics that are very relevant to our purpose.

Intuitively, a property is intrinsic if any object exemplifies it or can exemplify it independently of anything else in the Universe. A typical example of an intrinsic property is "having a mass". An extrinsic property like "having a weight" is instead attributable to a body only in relation with something else and in fact our weight on the Moon is different from our weight on Earth, while our mass is identical in the two environments. We could also characterize the distinction by using a modal language: "If something has an intrinsic property, then so does any perfect duplicate of that thing; whereas duplicates situated in different surroundings will differ in their extrinsic properties. (Lewis 1983: 197). A duplicate of an object having the same mass will different in its extrinsic properties (weight) if located in different environments. In Bohr's language, this translates into the claim that if the *same* quantum system S (two duplicates) is situated into two different measurements settings ("surroundings"), it will display different extrinsic properties and in this sense these properties are not intrinsically possessed, but purely extrinsic.

However, note that there is an important difference between an extrinsic (or relational) property and a contextual one.[6] "Being a brother of" is relational because it cannot be exemplified independently of a sister or a brother (same for having a certain weight that cannot be instantiated without a massive object) but each sibling has intrinsic properties like mass, volume, shape, genetic code etc. that do not depend on what else is around. If parents, ancestors and siblings were to disappear, all of the properties of one of the two siblings would become non-relational in the sense that relational properties of this kind seem to be reducible to the intrinsic properties of the *relata*.

A contextual property P, instead, has features that "depend much more" on the surroundings, in the sense that P is *literally unknowable* without an interaction with the environment that in principle alters it. An extrinsically possessed set of properties (having a certain weight) can be measured only given a certain setting, but is predefinite before a measurement. This situation is rather unlike contextualism as formally established by Kochen and Specker's theorem, and which, as mentioned above, is implicit in the experimental practice and in conceptual analysis and consequently explicitly present in Bohr's philosophy. Bohr was the first to realize that *the contextual properties P of a system S are literally unknowable without a measurement interaction that however alters them*, a revolutionary conclusion of quantum theory that must be accepted independently of its different interpretations and formulations.

One can insist on the fact that the consequence of the contextuality principle are not so dramatic: as the bohemian physicist Goldstein often remarks, it is quite trivial to recognize that the outcome of an experiment depends on the experiment context that is used to perform it: "experiments differ and different experiments usually have different results. The misleading reference to measurement, *which suggests that a pre-existing value of A is being revealed*, makes contextualism seem more than it is" (Goldstein 2017, my emphasis).

However, both Heisenberg indeterminacy relations and Bohr' holistic philosophy of quantum mechanics can be stated in a more general way without any reference to the *changing measurement settings*, since they are a consequence of the inescapable fact that also Goldstein seems to downplay at least in this passage, namely that "*there are no measurement apparatuses that allow the determination of the quantum state of an object*" (Allori and Zanghì 2006, p.254, my translation, my emphasis). By paraphrasing the words of these two Italian scholars, and drawing their consequence for the purpose of this paper, *this is the true impassable epistemic limit imposed by the theory, which in its turn depends on the smallness of quantum systems and of the quantum of action*.

---

[6] There are subtleties about the difference between relational and extrinsic properties that need detail us (see Marshall and Weatherson 2018)

In a word, there is a sense in which the intrinsic magnitudes of any quantum system S are unknowable, exactly because measurements don't reveal preexisting properties, something that has been appropriately endorsed also by Bohr's archenemy, namely John Bell: "the word [measurement] very strongly suggest the ascertaining of some pre-existing property…Quantum experiments are not just like that, *as we learnt especially from Bohr*. The results have to be regarded as the joint product of "system" and "apparatus" (quoted in Whitaker 1989, 180, my emphasis).

Evidence for his full endorsement of the *holistic nature of any measurement interaction* is provided by Bohr himself: "While, within the scope of classical physics, the interaction between object and apparatus can be neglected or, if necessary, compensated for, in quantum physics this interaction forms an *inseparable* part of the phenomenon." (Bohr 1963, 4, my emphasis). On the basis of this and other passages (see Howard 1994), Bohr is here referring with other words to the typically quantum mechanical phenomenon of non-separability. This implies that *two possible readings* of the expressions "joint product" and "inseparable part of the phenomena" - and therefore of the nature of the ensuing epistemic limit dictated by quantum mechanics - must be excluded.

The first reading to be ruled has it that this "joint product" can be interpreted as a mere neopositivistic appeal to the fact that it is *meaningless* to talk about state-dependent properties of quantum entities independently of a measurement apparatus.[7] The fact that one cannot *know* the properties of a system without creating an inseparable whole with the measuring apparatus (an epistemic constraint) is *not* equivalent to claiming that it is meaningless to refer to a *ontically* indeterminate, because not-yet measured and therefore not-existent properties of quantum systems (see above). An extra reason to reject this first reading of Bohr's expression is that the neopositivist conception of the meaning of a sentence has been substantially abandoned. One must distinguish ontological claims from epistemic claims: before Wiles' solution of Fermat last theorem, the proof lacked, but now we discovered that it is true, which seems to imply that it was nonetheless true before the actual demonstration, that is, before we came to know that it is true. One would not want to claim that before the actual proof its statement was meaningless!

---

[7] For this reading see Redhead 1987, 49-51, and Beller and Fine 1994.

The second reading to be rejected involves the disturbance view of measurements interactions, that Bohr plausibly abandoned already as a consequence of his reply to EPR (Bohr (1935)[8]. In this reply in fact, Bohr wrote quite clearly that "there is in a case like that just considered no question of a *mechanical disturbance* of the system under investigation during the last critical stage of the measuring procedure (Bohr, 1935 p. 700, my emphasis) [9] The italicized expression means that Bohr rejected non-local effects, exactly like Einstein What is the meaning, therefore, of Bohr's holism"? In order to answer this question, which is crucial for the purpose of this paper, we must quote Bohr in full and try to clarify his somewhat obscure prose.  the view that the measurement outcomes are "the joint product of systems and apparatuses" …implies that any form of knowledge of *the single quantum system is impossible*, since it depends on the influence "of the very conditions which define the possible types of predictions regarding the future behavior of the system. Since these conditions constitute an inherent element of the description of any phenomenon to which the term "physical reality" can be properly attached, we see that the argumentation of the mentioned authors does not justify their conclusion that quantum-mechanical description is essentially incomplete" (ibid.). The non-separability of system and instrument implies that the properties of the system strictly depend on the kind of measurement we want to perform and cannot be described with the context provided by measurement apparatus.

In Kantian language, Bohr's infelicitous term "influences", which seems to suggest some sort of causal dependence, and to which he is referring to are the conditions of possibility of performing a quantum experiment (see Murdoch 1985), which, more faithfully to text, involve the principle of complementarity.

### 3    Complementarity and dispositionalism in Bohr's philosophy of quantum mechanics

The main claim that we want to defend in this section is that a dispositional reading of Bohr's philosophy of quantum theory is *one* clear way to formulate the problem of the existence of quantum

---

[8] By referring to EPR, in his 1949 Bohr wrote "there is in a case like that just considered no question of a mechanical disturbance of the system under investigation during the last critical stage of the measuring procedure (Bohr 1949, 235). The "disturbance argument" was suggested by Heisenberg in 1930, in his optical, "microscope argument

[9] The hypothesis that Bohr abandoned the disturbance view according to which the quantum system has a previous definite value which is unknowable because its state is disturbed by the measurement had already been attacked by (Bohm 1951, Folse 1985, Faye 1991, Whitaker 2004, 1324) but 235). The "disturbance argument" was suggested by Heisenberg in 1930, in his optical, "microscope argument

limits of knowledge as he saw them (for this claim see Dorato 2007). A disposition is a quadruple constituted by a quantum object Q possessing the disposition D, a stimulus S for the manifestation of the disposition and the manifestation event E. In symbols <O, D, S, E>. Typically, macroscopic dispositions like fragility are in principle reducible to the molecular structure of the glass. Analogously for flammability, or even for mental dispositions like irritability, etc. The main idea that we want to illustrate is that quantum systems *S* possess *irreducible dispositions D* whose manifestations depend on *different, incompatible measurement contexts* that are the stimula *W* of the manifestation *event E*, where *E* is simply the measurement outcome.

Let me say at the outset that there is *no* explicit evidence in Bohr's published and unpublished texts of his endorsement of a dispositional interpretation of quantum mechanics. And yet I think that such re-interpretation does not distort Bohr's thought too much, that it can clarify some of the most obscures points of his doctrine and is particular useful to illustrate one way of formulating the issue of the quantum limit of knowledge that is the aim of the current paper.

In order to present this account and its correlated propensity-like approach to probability is opportune to recapitulate well known-doctrines expressed by Bohr already in the Como lecture (Bohr 1928). First of all, Bohr claims that doing science presupposes a freedom of choice on the part of the experimenter. Therefore, the free decision to find out about the *spatiotemporal aspect* of the single dispositional nature of a quantum system O, implies the decision to have it interact with a classical, macroscopic apparatus (the *stimulus* S for the manifestation of the disposition possessed by O) that is appropriate to reveal it by making it *manifest* in the outcome (the event E). If, on the other hand, we want to reveal what he called the "causal" aspect (the conservation principles) of the single disposition possessed by the same system O, we must set the stimulus for its manifestation (the measurement apparatus) in the appropriated way. Since the two measurement settings are not only different but non-realizable at the same time, the very same system's spatiotemporal aspect of its disposition is lost.

To apply these hypotheses to a concrete case study, think of his 1927 famous discussion with Einstein at the 5[th] Solvay conference. If we want to calculate the impact of an electron on the movable screen imagined by Einstein to be suspended on the top of the experiment table with a spring, we could in principle be able to measure the exchange of momentum between the electron and the screen in virtue of the conservation of momentum, which Bohr referred to as the "causal principle".[10] But this in its turn requires a very precise measurement of the velocity of the screen and therefore, in virtue of Heisenberg's indeterminacy relations, an indeterminacy about the position of the pointer of the first screen. This indeterminacy that will cause a loss of the interference on the second screen, the typical behavior of waves. Note that in his objection to Einstein's thought experiment, Bohr treats a macroscopic system as being subject to the quantum indeterminacy relations and equates it, with a "contextualist" move, (Zinkernagel 1916), with a quantum system!

---

[10] In virtue of the presence of a second two-slits screen posed behind the mobile one, according to Einstein one could at the same time determine in which of the two slits the particle went through. Position and momentum would be simultaneously determined. It is controversial whether Einstein's thought experiment was really targeting Heisenberg.

To the extent that such relations are an essential mark of the quantum description of reality, Bohr typically referred to classical systems that are *not* subject to the relations for as a necessary condition for the possibility of any quantum measurement. In his response to Einstein's challenge, however, a macroscopic system is untypically regarded as *quantum* (i.e. *not* subject to Heisenberg's relations) depending on the measurement context treated as non-separable from the impacting, "standard" or "real" quantum system (a particle). There is a (weak) sense in which his holism overcomes this difficulty: the complementarity principle tells us that *if* the first screen is solidly attached to the table, the position of the particle going through the slit (the spatiotemporal aspect of the *single* disposition) is manifested and therefore becomes known, but then the exchanged moment cannot be calculated because the corresponding manifestation cannot occur.[11] On the other hand, the presence of the mobile screen is an appropriate stimulus for the manifestation of the momentum of the incoming particle, but the complementary spatiotemporal aspect of the disposition is lost, because we cannot determine in which slit of the second screen the particle has gone through.

As a matter of fact, additional evidence for the appropriateness of our dispositionalist interpretation of Bohr's philosophy can be obtained by discussing the two-slits experiment, which, also according to Feynman contains all the mysteries of quantum mechanics. In Bohr's language, the wave-like behavior of quantum systems depends on the fact that the stimulus of the disposition is such that *both* slits are open (so as to generate interference). The particle-like "aspect" of the system that is revealed after it enters both slits is such that only a small location on a spot of the second screen can be seen, because the second screen used to detect the arrival of the electron (the complementary stimulus for the manifestation of the disposition) is built to localize the wave in a certain point via a nonlocal phenomenon in which a spatially extended way collapses to a point of the screen.

---

[11] For an attentive reconstruction of the important episode in the history of quantum mechanics, see Bacciagaluppi and Valentini (2009).

The quantum limit of knowledge in the interpretation we are proposing can therefore be redescribed in terms of the impossibility of coming to know at the same time the two manifestations E of the single quantum dispositions within the same holistic measurement context (stimulus). Recall that according to Bohr, two properties are complementary if and only if they are *mutually exclusive* and *jointly exhaustive*. We say that they are *mutual exclusive* because, by being necessitated to use *one incompatible stimulus* S at the time (measurement context), obtaining one outcome E (one of the two possible manifestations of the disposition) excludes the possibility to come to know at the same time the other possible manifestation of the single, intrinsic disposition possessed by the quantum system. As we know, complementary properties cannot be simultaneously revealed by the *same* experiment, given that any apparatus either selects spatiotemporal features of the quantum properties or keeps track of the exchange of momentum. On the other hand, if we refer to a quantum system *before* measurement, the complementary properties that can be manifested by the disposition of the quantum system must be regarded as *jointly exhaustive*: any attempt at attributing a not-yet measured system only *one* of the two possible manifestation or property would yield an incomplete description. *In fact, we can attribute the unmeasured electron an intrinsic, single disposition that in itself is unknowable but that is capable to manifest incompatible outcomes according to incompatible measuring circumstances.*

In a word, the whole experimental set-up provides the context of manifestation of quantum mechanical properties, that therefore must be conceived as being essentially dispositional. The probabilistic character of the quantum mechanical dispositions renders the language of single cases propensity particularly apt for giving an account of the origin of frequencies when many experiences are repeated. Furthermore, by giving us a chance to reinterpret the relationship between the principle of complementarity of non-commuting observables, the language of dispositions also helps us to make sense of what we find in experimental practice independently of the chosen interpretation of quantum mechanics: what a quantum entity *is* contextually depends not just on what other entities it has interacted with in the past (non-locality), but also on the whole experimental arrangement with which it interacts (quantum holism, which also explain why Bohr invoked classical objects as being subject to the quantum realm).

## 4 Rovelli's relationism and dispositionalism about quantum mechanics

The theme of the relativization of physical quantities that were previously regarded as absolute goes through the history of physics as a leading thread. With the Galilean principle of relativity, for example, we discovered that the notion of velocity is relative to an arbitrarily chosen inertial frame. Analogously, within special relativity, "space by itself and time by itself, are doomed to fade away like mere shadows and only a kind of union between the two will preserve an independent reality" (Minkowki 1908).[12] However, this process of relativization was accompanied by the search for *invariant* elements, in this case, of relativity the four-dimensional metric. In Rovelli's relational quantum mechanics, the relativization is even more radical as it extends to the *identity of quantum-physical systems*, at least to the extent that (i) the possession of some intrinsic properties is essential to the identity of an object[13] and (ii) no entity can exist if it does not have an intrinsic identity. The identity of objects or physical systems S in the relational quantum world envisaged by Rovelli is therefore purely structural, in a ontic structural realism:[14] post-measurement assertions like "relative to observer *O*, system S has value q" are true only relative to O. For another observer *P*, S could have no definite value until P interacts with O+S, and, as we are about to see, the outcome that P attributes to O+S and therefore to O's observation may differ from q. In the special theory of relativity, on the contrary, the fact that "body B has length L relative to system S" holds for any possible observer and there is in any case the Minkowski metric that plays the role of an invariant element. In relational quantum mechanics, Rovelli seems to have no new absolute quantity to propose, except, perhaps the probability rules (the Born rules) associated to all measurement interactions.

After the necessary qualification that the term "observer" in Rovelli's approach refers to any quantum system whatsoever (not just to conscious beings), it is important to add that the so-called be-ables (by Bell contrasted to observa-bles) of his interpretation are the outcomes of the interactions between two quantum systems, namely *events*, which are the building blocks of spacetime. Each of these events in succession is a local perspective on the rest of the universe, or a local, non-necessarily continuous worldline W whose temporal order cannot be extended to any other systems.[15]

---

[12] It is more controversial whether in general relativity all kind of motion is relative: Malament (2000) proved a theorem in which rotation has to be considered an intrinsic property.

[13] Philosophers also talk about irreducible elements that provide identity to objects (heacceitates) independently of their properties

[14] For an overview of the rich literature on ontic structural realism, see French 2014.

In my previous reconstruction of Rovelli's relational approach (Dorato 2016), I suggested that it is preferable to defend a dispositionalist interpretation of the indeterminacy of any quantum system before a measurement interaction, so that both non-interacting quantum systems S and "observers" O have no intrinsic properties, except dispositional ones. In other words, such systems S have intrinsic dispositions to correlate with other systems/observers $O$, which manifest themselves as the possession of definite properties $q$ relative to those $O$s. The real events of the world are the 'realization' (the 'coming to reality', the '*actualization*') of the values $q$, $q_0$, $q_1$, . . . in the course of the interaction between physical systems.

Exactly like Wigner, Rovelli reconstructs Von Neumann dualistic evolution by distinguishing between observers that are *internal* to the interaction $S+O$ and those that are *external* to it (in the example above, the observer $P$). Wigner's friend $O$ inside the room is internal to the interaction with $S$ and perceives a definite event as a result of the correlation, while Wigner, that is external to the interaction, consider the system S+O in a superposed state until he correlates with the joint system: "In one word: value assignment in a measurement is not inconsistent with unitary evolution of the apparatus+system ensemble, because value assignment refers to the properties of the system with respect to the apparatus, while the unitary evolution refers to properties with respect to an external system." (Rovelli 1998, 19).

We can now present three possible senses in which Rovelli's relational quantum mechanics poses a in-principle limitation to the our knowledge of the physical world. The first sense is given by Rovelli's identification between the interaction between any two systems and an *exchange of information*. It then follows a priori that no system can have information about itself: "*The unitary evolution does not break down for mysterious physical quantum jumps, due to unknown effects, but simply because O is not giving a full dynamical description of the interaction. O cannot have a full description of the interaction of S with himself (O), because his information is correlation, and there is no meaning in being correlated with oneself.* (Rovelli 1997, p. 205, emphasis in original).

The second sense referred to above is even more important from the viewpoint of this paper since in Rovelli's interpretation the limits of knowledge imposed by quantum theory extend not just to the dispositional properties of single, non-measured quantum systems but also to *a temporal succession* of measurement outcomes (events). In other words, the worldlines that are invariant for all observers and that typically constitute the relativistic four-dimensional world according to Rovelli would be described differently by different observers: "in quantum mechanics different observers may give different accounts of the same sequence of events" (Rovelli 1996, p. 4). However, how can sequences of events retain their identity if the outcomes that different observers measure *are* events and events have identities that depend on the interaction with different observers? Doesn't it follow that in relational quantum mechanics not even the sequence of events is objectively knowable since it is not invariant across different observers?

---

[15] Interestingly, in Rovelli' interpretation this non-extendibility is forbidden not just for relativistic reasons. Stress on the discontinuous character of Rovelli's ontology is defended in Laudisa and Rovelli (2019).

The following example, discussed by Brown, will help to shed some light on this difficult case. With obvious notation, suppose that at time $t_1$ the state of the quantum system $S$ is in the superposition:

$$|\Psi_S\rangle = \alpha |\uparrow\rangle_S + \beta |\downarrow\rangle_S, \qquad \text{with } |\alpha|^2 + |\beta|^2 = 1 \tag{1}$$

Suppose that at time $t_2$ $O$ measures $S$ and that, relative to her (or it), the outcome is $|\uparrow\rangle_S$ According to the relational interpretation, the state of $S$ for $O$ evolves from $|ready\rangle_O |\Psi\rangle_S$ at time $t_1$ to $|\Psi\rangle_{S/O} = |\uparrow\rangle_S$ at time $t_2$. The index $S/O$ expresses the fact that the outcome "spin up" revealed by the interaction between the system S and the observer O is relative to the latter. Given the linearity of the evolution of the $\Psi$ function, according to another observer $P$ who at time $t_2$ has not yet interacted with $S+O$, the two systems $S+O$ at time $t_2$ are still superposed, so that relative to $P$ at time $t_2$ we have

$$|\psi>_{SO/P} = \alpha |up>_O |\uparrow>_S + \beta |down>_O |\downarrow>_S. \tag{2}$$

Let us quote Brown in full: "the probability that $P$ will find the state at [a later time] $t_3$ to be $|up>_O |\uparrow>_S$ …. is $|\alpha|^2$, and the probability of $|down>_O |\downarrow>_S$ is $|\beta|^2$. So, as von Neumann taught us, the *probabilities* agree. But notice: if we are to take relational quantum mechanics seriously, *nothing* said so far prevents it from being the case that P finds $|down>_O |\downarrow>_S$ at $t^3$, and thus S being spin-down for P, even though S was spin-up for O!" (Brown 2009, p. 690).

Is this a contradiction? The sequence of events: '$O$ ready to measure $S$ (1) at $t_1$, then '$S$ finds spin up at $t_2$', then 'P finds that S found spin down at $t_3$' is different from the sequence that began with $O$ ready to measure and then led to the observation that according to $S$ is spin down, because *the correlation between S and O is described differently from O and P*. It is this correlation that is the objective selfsame ' meta-event' generating non-invariant measurement events when systems and observers interact. By distinguishing between processes of correlations $P$ from processes of the outcomes of these correlation the temporal evolution remains objective at the metalevel and does not conflict with the constraints of relativity, despite the discontinuous character of the worldline interactions.

The third of the senses mentioned above shows how Rovelli' relationism shows important similarities with Everettian quantum mechanics whenever an observer is the target of an another observer. For simplicity suppose that I am *O* and you are *P*. For me *S* has spin up and I *observed* that *S* has spin un, for you, after reduction of (2), *my* interaction with *S* resulted in spin down and I observed spin down. However, this relativization won't do: how can I observe that *S* has spin up (for me) and at the same time observe that it has spin down (relative to you?). Our observations cannot be relativized! Can I observe and not observe that the same system *S* has spin up and not spin up? In order to avoid the contradiction some sort of Everettian quantum mechanics is called for, in such a way that relative to a state of the system, I observe that *S*'s spin is up and you observe that I observe that spin is up while relative to the other state of the system I observe that the spin is down and you observe that I observe that spin is down. Coherence is then reestablished.

The reader can establish by herself the cogency of this claim by reading the following long quotation: "There does not, in general, exist anything like a single state for one subsystem of a composite system. Subsystems do not possess states that are independent of the states of the remainder of the system, so that the subsystem states are generally correlated with one another. One can arbitrarily choose a state for one subsystem, and be led to the relative state for the remainder. Thus we are faced with a fundamental relativity of states, which is implied by the formalism of composite systems. It is meaningless to ask the absolute state of a subsystem—one can only ask the state relative to a given state of the remainder of the subsystem. (Everett 1956, 103; 1957, 180).

5  Rovelli's relational quantum mechanics and the universal state function

In sum, in relational quantum mechanics there are no absolute states with definite properties. However, if such states ought to be interpreted as suggested above as being irreducibly dispositional, ontologically there must be a purely dispositional, quantum state of the universe, so that quantum cosmology is a legitimate enterprise. From an epistemological point of view, however, quantum cosmology requires that one studies large segments of the universe *in relation to other segments*. Recall that in relational quantum mechanics there is no fact of the matter about whether two different observers *O* and *P* get the same result out of an interaction with the system *S*, since this is a question about their absolute states.

It follows that if *S* is the whole universe, then, qua isolated system, also S is an absolute system with potentially infinite, purely dispositional properties that, lacking any external stimulus, cannot be manifested. Nothing can interact with the quantum universe *S* in principle, for there is no external observer. Consequently, the quantum universe S can be only *partially known* by interacting with parts of it *from within*, namely by dividing it into two parts, one of which, the observer *O*, must be contained in *S*. If the quantum universe can be described only from within, we must somehow consider all the possible compatible perspectives about it, each of which depends on a cut of the universe into two parts, a system and an observer and the way the exchange information. The consequence of this is of momentous importance: *Rovelli's relational approach to quantum mechanics entails that the quantum state of the universe cannot be know!* Another, more dramatic instance of the impossibility of self-measurement of a quantum system!

References

Allori V. and Zanghì N. (2005), "Un viaggio nel mondo quantistico", in Allori V., Laudisa F. and Zanghì (eds.) *La natura delle cose.* Roma, Carocci, pp.229-390.

Bell, John S., 1964, "On the Einstein-Podolsky-Rosen Paradox", *Physics*, 1(3): 195–200

Beller, Mara, and Arthur Fine. 1994. "Bohr's Response to EPR." In *Niels Bohr and Contemporary Philosophy*: *Boston Studies in the Philosophy of Science, Vol. 153*. Edited by Jan Faye and Henry Folse, 1-31. Berlin: Springer.

Bohr, Niels (1928), "The Quantum Postulate and the recent Development of atomic theory", Nature, 121 Supp. 1928

Bohr, Niels (1935), Can quantum mechanics descriptions of physical reality be considered complete? in Physical Review, 48, 696-702

Bohr, Niels. 1949. "Discussion with Einstein on epistemological problems in atomic physics." In *Albert Einstein: Philosopher−Scientist.* Edited by Paul A. Schilpp, 200−41. Evanston: The Library of Living Philosophers., Reprinted in N. Bohr. 1958. *Atomic Physics and Human Knowledge,* 32-66.

**Breuer, T., 1993, "The impossibility of accurate state self-measurements",** *Philosophy of Science*, 62: 197-214.

Brown M. (2009) "Relational quantum mechanics and the determinacy problem", in *Brit. J. Phil. Sci.* **6**, 679–695.

Clifton R. and Pagonis C. (1995), 'Unremarkable Contextualism: Dispositions in the Bohm theory', *Foundations of Physics*, 25, 2 (1995): 283.

Cushing, James (1994), *Quantum Mechanics: Historical Contingency and the Copenhagen Hegemony*, Chicago University Press.

Dalla Chiara M. (1977), Logical self-reference: set theoretical paradoxes and the measurement problem in quantum mechanics, *Journal of Philosophical Logic*, 6, pp. 331-347.

Dorato, M. (2007). "Dispositions, relational properties and the quantum world." In *Dispositions and Causal Powers*, edited by Max Kistler and Bruno Gnassonou , 249-270. Oldercroft: Ashgate.

Dorato M. (2016), "Rovelli's Relational Quantum Mechanics, Anti-Monism, and Quantum Becoming", in *The Metaphysics of Relations*, Anna Marmodoro and David Yates (eds.), Oxford: Oxford University Press, 235–262doi:10.1093/acprof:oso/9780198735878.003.0014

Everett, H., 1956, "The Theory of the Universal Wave Function". First printed in DeWitt and Graham (1973), 3–140. Reprinted as cited here in Barrett and Byrne (2012) 72–172.

Everett H. (1957), "Relative state formulation of quantum mechanics", in Barrett, J. and P. Byrne (eds.), 2012, *The Everett Interpretation of Quantum Mechanics: Collected Works 1955–1980 with Commentary*, Princeton: Princeton University Press. Bell, J. S., 1966, "On the Problem of Hidden Variables in Quantum Mechanics", *Review of Modern Physics*, 38: 447–52.

Forman, P. (1984), "Kausalität, Anschaulichkeit, and Individualität, or How Cultural Values Prescribed the Character and Lessons Ascribed to Quantum Mechanics," in *Society and Knowledge*, eds. Nico Stehr and Volker Meja. Transaction Books, 1984: pp 333-347.

French, Steven (2014), *The Structure of the World*, *Metaphysics and Representation.* OUP, Oxford.

Ghirardi G.C. (1985) *Sneaking a look at God's cards* Princeton University Press, Princeton, transl from *Un'occhiata alle carte di Dio*, Milano 1981

Gleason, Andrew M., 1957, "Measures on the Closed Subspaces of a Hilbert Space", *Indian University Mathematics Journal*, 6(4): 885–893. doi:10.1512/iumj.1957.6.56050.

Goldstein, Sheldon, "Bohmian Mechanics", *The Stanford Encyclopedia of Philosophy* (Summer 2017 Edition), Edward N. Zalta (ed.), URL = <https://plato.stanford.edu/archives/sum2017/entries/qm-bohm/>.

Hawking S. (1981), "Is the end in sight for theoretical physics?", *Physics Bulletin*, Volume 32, Number 1, p.15-17.

Howard, Don. 1997. "A Peek behind the Veil of Maya." In *The Cosmos of Science.* Edited by John Earman and John Norton, 87-152. Pittsburgh: University of Pittsburgh Press..

Howard, Don 1994 "What Makes a Classical Concept Classical? Toward a Reconstruction of Niels Bohr's Philosophy of Physics", in *Niels Bohr and Contemporary Philosophy: Boston Studies in the Philosophy of Science 158.* Edited by Jan Faye and Henry Folse, 201–229., Dordrecht: Kluwer Academic.

Howard, Don (2004) Who invented the "Copenhagen Interpretation?" A study in mythology" *Philosophy of Science,* 71, pp.669-682.

Hughes R. (1989), *The Structure and Interpretation of Quantum mechanics*. Harvard University Press.

Isham C. (1995), *Lectures on Quantum Theory*, Imperial College Press, London.

Kochen, S. Specher E. (1967), "The problem of hidden variables in quantum mechanics", *Journal of Mathematics and Mechanics*, 17, pp. 59-87.

Laudisa, Federico and Rovelli, Carlo, "Relational Quantum Mechanics", *The Stanford Encyclopedia of Philosophy* (Winter 2019 Edition), Edward N. Zalta (ed.), forthcoming URL = <https://plato.stanford.edu/archives/win2019/entries/qm-relational/>.

Laudisa F. (2014), Against the no-go philosophy of quantum mechanics, *European Journal for Philosophy of Science*, volume 4, pages 1–17

Lewis, David, 1983, "Extrinsic Properties", *Philosophical Studies*, 44: 197–200.

Malament, David 2000: 'A No-Go Theorem about Rotation in General Relativity', In Festschrift for Howard Stein, edited by David Malament, Open Court Press

Marshall, Dan and Weatherson, Brian, (2018), "Intrinsic vs. Extrinsic Properties", *The Stanford Encyclopedia of Philosophy* (Spring 2018 Edition), Edward N. Zalta (ed.), URL = <https://plato.stanford.edu/archives/spr2018/entries/intrinsic-extrinsic/>.

Minkowski, H. (1908). "Space and time." In Einstein, *et al*. (1952), pp. 75–91.

Redhead M. (1987), *Incompleteness, Nonlocality and Realism,* Clarendon Press, Oxford.

Rovelli C., (1996), "Relational Quantum Mechanics*", International Journal of Theoretical Physics*, 35 1637.

Rovelli C. (1998), http://arxiv.org/pdf/quant-ph/9609002v2.pdf

Von Neumann J. (1955) *Mathematical Foundations of Quantum Mechanics*, Princeton University Press, Princeton (original edition 1932).

Whitaker, Andrew. 1989. *Einstein Born and the Quantum Dilemma*, Cambridge: Cambridge University Press.

Zinkernagel, Henrik. 2016. "Niels Bohr on the wave function and the classical/quantum divide." *Studies in History and Philosophy of Modern Physics* 53, 9–19.